\documentclass[3p,sort&compress]{elsarticle}

\usepackage{lineno,hyperref}
\modulolinenumbers[5]
\usepackage{amssymb}
\usepackage{amsmath}
\usepackage{epsfig} 
\usepackage{multirow}
\usepackage{adjustbox}
\usepackage{threeparttable}
\usepackage{colortbl}

\journal{Nuclear Physics A}

\def\8B{$^8$B}
\def\7Be{$^7$Be}
\def\Rm{$R_{\rm m}$}
\def\Rp{$R_{\rm p}$}
\def\dsdt{d$\sigma/$d$t$}
\def\TR{$T_{\rm R}$}
\def\vq{\textit {\textbf q}}
\def\vr{\textit {\textbf r}}
\def\vs{\textit {\textbf s}}
\def\vb{\textit {\textbf b}}

\begin{document}

\begin{frontmatter}

\title{Nuclear-matter distribution in the proton-rich nuclei $^7$Be and $^8$B from intermediate energy proton elastic scattering in inverse kinematics}

\author[address1]{A.V.~Dobrovolsky\corref{mycorrespondingauthor}}
\cortext[mycorrespondingauthor]{Corresponding author}
\ead{Dobrovolsky\_AV@pnpi.nrcki.ru}

\author[address1]{G.A.~Korolev}
\author[address1]{A.G.~ Inglessi}
\author[address1]{G.D.~Alkhazov}
\author[address2]{G.~Col\`{o}}
\author[address3]{I.~Dillmann}
\author[address3]{P.~Egelhof}
\author[address3]{A.~Estrad\'{e}}
\author[address3]{F.~Farinon}
\author[address3]{H.~Geissel}
\author[address3]{S.~Ilieva}
\author[address3]{Y.~Ke}
\author[address1]{A.V.~Khanzadeev}
\author[address3]{O.A.~Kiselev}
\author[address3]{J.~Kurcewicz}
\author[address3]{X.C.~Le}
\author[address3]{Yu.A.~Litvinov}
\author[address1]{G.E.~Petrov}
\author[address3]{A.~Prochazka}
\author[address3]{C.~Scheidenberger}
\author[address1]{L.O.~Sergeev}
\author[address3]{H.~Simon}
\author[address3]{M.~Takechi}
\author[address3]{S.~Tang}
\author[address3]{V.~Volkov}
\author[address1]{A.A.~Vorobyov}
\author[address3]{H.~Weick}
\author[address1]{V.I.~Yatsoura}

\address[address1]{Petersburg Nuclear Physics Institute, National Research Centre Kurchatov Institute, Gatchina, 188300 Russia}
\address[address2]{Dipartimento di Fisica, Universit\`{a} degli Studi di Milano and INFN, Sezione di Milano, Via Celoria 16, 20133 Milano, Italy}
\address[address3]{GSI Helmholtzzentrum f\"{u}r Schwerionenforschung GmbH, 64291 Darmstadt, Germany}

\begin{abstract}
	Absolute differential cross sections for elastic $p^7$Be and $p^8$B small-angle scattering were measured in inverse kinematics at an energy of 0.7~GeV/u at GSI Darmstadt. The hydrogen-filled ionization chamber IKAR was used as an active target to detect the recoil protons. The projectile tracking and isotope identification were performed with multi-wire proportional chambers and scintillation detectors. The measured cross sections were analysed using the Glauber multiple-scattering theory. The root-mean-square (rms) nuclear matter radii $R_{\rm m} = 2.42 (4)$~fm for \7Be and $R_{\rm m} = 2.58 (6)$~fm for \8B were obtained. The radial density distribution deduced for \8B exhibits a proton halo structure with the rms halo radius $R_{\rm h} = 4.24 (25)$~fm. A comparison of the deduced experimental radii is displayed with existing experimental and theoretical data.
\end{abstract}


\end{frontmatter}


\section{Introduction}

The recent development of radioactive isotope beam techniques has opened excellent opportunities to study the structure of light unstable nuclei far from the valley of stability. Among others, new properties of the nuclear matter were discovered. The weak binding of the last bound nucleons may cause in special cases the formation of neutron (proton) skins on the surface of the nucleus and the formation of a halo structure~\cite{Tanihata13,Jonson04,Ozawa01,Tanihata96}. It has been found that the existence of a halo corresponds to a large extension of the matter density distribution beyond the nuclear core. The halo structure manifests itself by large interaction (reaction) cross sections, by increased removal nucleon cross sections and by narrow momentum distributions of reaction products in the processes of nuclear break-up and Coulomb dissociation. An increase in the root-mean-square (rms) radius \Rm ~of the radial distribution of the nuclear matter in nuclei near the neutron drip line was the first indication of a halo in exotic nuclei such as $^6$He, $^{11}$Li, $^{11}$Be and $^{14}$Be~\cite{Ozawa01,Tanihata96}. These investigations showed that most of the halo nuclei have neutron halos, while the formation of proton halos is much less probable due to the Coulomb barrier effect~\cite{Tanihata13}.

The proton drip-line nucleus \8B is considered to be one of the most interesting candidates for the occurrence of a proton halo since it has a very small proton separation energy $S_{\rm p} = 136.4$~keV~\cite{Wang12}. The study of the proton-rich nuclei \8B and \7Be (\7Be being the presumable core in \8B) is important for nuclear physics.
The \8B proton-halo problem has received much attention from both experimental and theoretical points of view. The halo structure of \8B was suggested for the first time by the Osaka group~\cite{Minamisono92} to explain the unusually large quadrupole moment $Q$ of this nucleus as compared to the value for the mirror nucleus $^8$Li. The experimental data for $Q$ were well reproduced by a wave function with a long proton tail obtained for \8B in the shell model using the Cohen-Kurath interaction. Howerer, theoretical investigations~\cite{Csoto93,Nakada94} have shown that the large quadrupole moment of \8B can be explained without the existence of a proton halo. At present, the main evidence for the halo structure of \8B is obtained from experiments on break-up reactions. The measurements of the momentum distribution of \7Be fragments from the \8B break-up reactions performed at GSI have shown a much narrower distribution than the one for stable nuclei~\cite{Schwab95}. Similar measurements have later been performed at different beam energies~\cite{Kelley96,Negoita96,Smedberg99,Cortina03}. Moreover, the one-proton removal cross section is enhanced as compared to the case of a nucleus with a tightly bound proton~\cite{Schwab95,Smedberg99,Cortina03,Blank97,Fukuda99,Warner04}, what is also a signature for an extended valence-proton wave function.

The size and the shape of the radial distribution of the nuclear matter are fundamental properties of nuclei and can be the most convincing evidence for the proton halo structure. The matter density distribution in \8B was determined and the rms matter radius \Rm ~was obtained in measurements of interaction or reaction cross sections~\cite{Negoita96,Fukuda99,Tanihata88,Warner95,AlKhali96,Obuti96,Fan15}. However, the values of \Rm ~deduced using different versions of the Glauber model are widely scattered, ranging from 2.38(2)~fm to 2.61(8)~fm. To reproduce reaction cross sections for \8B measured at low energies, Warner $et~al$.~\cite{Warner95} used microscopic calculations with a matter distribution obtained by the Osaka group~\cite{Minamisono92}. The nuclear properties of \8B were the subject of much theoretical work performed in the last years, see Refs.~\cite{Csoto93,Baye94,Brown96,Kitagawa99,Furutachi09,Henninger15,Varga95,Fayans95,Grigorenko98,Patra98,Chandel03,Wang09,Pastore13} and references therein.

Information on the structure of \7Be, the presumable core of \8B, is rather scarce. The matter radii $R_{\rm m} = 2.31 (2)$~fm~\cite{Tanihata88} and $R_{\rm m} = 2.36 (6)$~fm~\cite{Warner01} were obtained through measurements of the interaction and reaction cross sections. The proton radius \Rp = 2.507(17)~fm~\cite{Lu13} was derived from the charge radius $R_{\rm ch}$ measurement in a laser spectroscopy study~\cite{Nortershauser09}. The proton-rich \7Be nucleus is a weakly bound two-body system with a separation energy of 1.59 MeV for break-up into $^3$He and $^4$He. The structure of \7Be was discussed in many theoretical investigations~\cite{Csoto93,Negoita96,Varga95,Fayans95,Grigorenko98,Patra98,Chandel03,Wang09,Pastore13,Shen96,Krieger12,Carlson15}. A cluster structure of \7Be (as well as of the \8B nucleus) was supposed in Refs.~\cite{Csoto93,Varga95,Grigorenko98}. The large cross section ($\sigma$ = 242 mb) for the He break-up channel measured at GANIL~\cite{Negoita96} supported the concept of the cluster structure for \7Be.

Note that the study of the structures of \8B and \7Be is important also for nuclear astrophysics. These nuclei play an essential role in the solar neutrino production. The \8B nucleus is produced in the sun through the $^7$Be($p,\gamma$)$^8$B reaction and emits a high energy neutrino, which can be detected in terrestrial experiments~\cite{Adelberger11}. The proton capture rate in \7Be strongly depends on the \8B structure. The size of \8B and the shape of the proton density distribution at large distances determine the proton capture rate and may be used in theoretical calculations of the solar neutrino flux~\cite{Adelberger11,Riisager93,Csoto98}.

The objective of the present work was to obtain in a single experiment information on the structure of \8B and its supposed core nucleus \7Be in order to derive a convincing conclusion about the existence of the halo in \8B. The proton-nucleus elastic scattering at intermediate energies around 700--1000 MeV is considered as one of the best methods to determine matter density distributions in stable nuclei~\cite{Tanihata13,Alk78,Sakaguchi17}. In order to study exotic nuclei, experiments in inverse kinematics using radioactive nuclear beams and the active target IKAR, a hydrogen-filled time projection ionization chamber, were proposed and then performed in a first experiment~\cite{Alk92,Alk97}. It turned out that small angle proton scattering is particularly sensitive to the halo structure~\cite{Alk92}. Therefore, in order to study the spatial structure of halo nuclei it is important to measure with high accuracy the absolute differential cross sections for proton elastic scattering at small momentum transfers. An analysis of the shape of the measured cross sections makes it possible to determine the nuclear matter distributions and radii of the nuclear cores and halos~\cite{Alk92,Alk02}. In further experiments performed at GSI Darmstadt by the PNPI--GSI Collaboration the method was successfully applied to study the stable $^4$He, $^6$Li and the neutron-rich isotopes $^6$He, $^8$He, $^8$Li, $^9$Li, $^{11}$Li, $^{12}$Be and $^{14}$Be~\cite{Alk02,Neumaier02,Egelhof02,Dobrov06,Ilieva12}.

The present paper reports new results obtained for the proton-rich isotopes \7Be and \8B. The first results on the study of the \8B structure in comparison with its mirror nucleus $^8$Li were briefly presented in a Letter published in Ref.~\cite{Korolev18}.

\section{Experiment}

The experiment was performed at the GSI Darmstadt. A primary $^{22}$Ne beam produced by the UNILAC--SIS accelerator complex was focused on an 8~g/cm$^2$ Be production target at the entrance of the FRagment Separator (FRS). The produced beryllium and boron ions were separated according to their magnetic rigidity and to their nuclear charge by inserting an achromatic (2.7 g/cm$^2$) aluminium degrader at the dispersive central focal plane of the FRS. The energy of the secondary beam at the centre of the hydrogen target was 701~MeV/u for \7Be and 702~MeV/u for \8B with an energy spread of 1.3\%. The mean energies of the beam particles were determined with an accuracy of about 0.1\%. The beam intensity was $\sim 3 \cdot 10^3$ s$^{-1}$. The contamination from other nuclei was below the 0.1\% level.

\begin{figure} 
	\centering
	\psfig{file=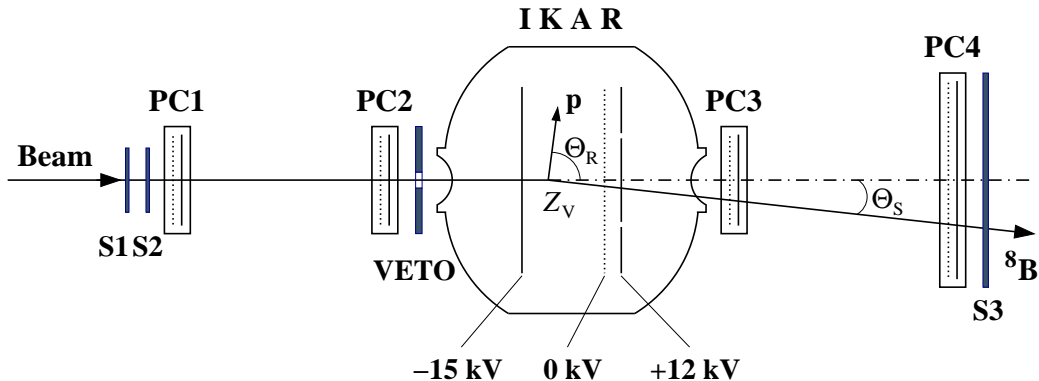,scale=0.9} 
	\caption{Schematic view of the experimental set-up for small-angle proton elastic scattering on exotic nuclei in inverse kinematics. The hydrogen-filled ionization chamber IKAR serves simultaneously as a gas target and a detector for recoil protons. For the sake of simplicity only one chamber module of six identical ones is shown. The tracking system consisting of four multi-wire proportional chambers PC1--PC4 determines the scattering angle $\varTheta_{\rm S}$ of the projectile. $\varTheta_{\rm R}$ is the recoil proton scattering angle and $Z_{\rm V}$ is the vertex point of the interaction. The scintillation counters S1--S3 and VETO are used for beam identification and triggering.}
	\label{setup}
\end{figure}

A schematic view of the experimental layout is displayed in Fig.~\ref{setup}.
The main part of the set-up was the ionization chamber IKAR, filled with pure hydrogen at a pressure of 10 bar, which served simultaneously as a gas target and a recoil proton detector.
The active target IKAR was developed at PNPI and was originally used in experiments on small-angle hadron elastic scattering~\cite{Vor74,Vor82,Burq78,Burq83}.
The chamber consists of six identical modules. Each module is an axial ionization time-projection chamber consisting of an anode subdivided into a central electrode and a concentric electrode, a cathode and a grid.
At the applied high voltages and for the used gas pressure the electron drift time from the cathode to the grid is 23 $\mu$s.
The signals from the electrodes provide the energy \TR~ of the recoil proton (or its energy loss in case it leaves the active volume),
the scattering angle $\varTheta_{\rm R}$ of the recoil proton and the vertex point $Z_{\rm V}$ of the interaction~\cite{Neumaier02}.

Thin $\alpha$-sources of $^{241}$Am were placed on the cathodes and grids, which permitted an energy calibration.
The procedure of energy calibration was described in detail in Ref.~\cite{Neumaier02}. The energy resolution of IKAR was $\sim$45--55 keV (sigma).
The $\alpha$-sources were used also for tracing a small correction which takes into account the limited transparency of the grid and the loss of drifting electrons through adhesion to electronegative impurities in the gas. These losses were continuously controlled by measuring the difference in the positions of two $\alpha$-peaks corresponding to $\alpha$-particles emitted from the sources deposited on the grid and on the cathode, respectively~\cite{Burq83}.
The recoil protons were registered in IKAR in coincidence with the scattered beam particles. For the measurement of the differential cross section \dsdt,~ the four-momentum transfer squared $-t$ could be determined either from the measured recoil energy \TR, or from the value of the scattering angle $\varTheta_{\rm S}$ of the projectiles, which was measured by a tracking detector system consisting of 2 pairs of two-dimensional multi-wire proportional chambers (PC1--PC2 and PC3--PC4) arranged upstream and downstream with respect to IKAR. A set of scintillation counters (S1, S2 and S3) was used for triggering and identification of the beam particles $via$ time-of-flight and d$E$/d$x$ measurements, while a circular-aperture scintillator VETO selected the projectiles which entered IKAR within an area with a diameter of 2 cm around the central axis. Cylinder bags filled with He gas (not shown in Fig.~\ref{setup}) were placed in-between each pair of multi-wire chambers in order to reduce the amount of multiple Coulomb scattering of the projectiles.

A detailed description of the experimental set-up and the procedure of the measurement is presented in Refs.~\cite{Neumaier02,Egelhof02,Dobrov06,Ilieva12}. The tracking of the projectiles was accomplished with the same system of multi-wire proportional chambers as in a previous experiment~\cite{Ilieva12}. The corresponding scattering angle $\varTheta_{\rm S}$ was obtained using the measured $x$ and $y$ coordinates in the multi-wire proportional chambers. The resolution for the scattering angle was determined by the position resolution and the angular spread due to multiple Coulomb scattering of the projectiles. The total angular resolution was estimated to be $\sigma_{\Theta}$ = 0.74~mrad for the case of \7Be, and $\sigma_{\Theta}$ = 1.00~mrad for \8B, as deduced from calibration measurements with unscattered beam particles.

\section{The data analysis}

The absolute differential cross section \dsdt~ was determined using the relation
\begin{equation}
\frac{{\rm d}\sigma}{{\rm d}t} = \frac{{\rm d}N}{{\rm d}t  B  n  \Delta L}~.
\end{equation}
Here, ${\rm d}N$  is the number of elastic proton-nucleus scattering events in the interval ${\rm d}t$ of the four-momentum transfer squared, $B$ is the corresponding number of beam particles impinging on the target, $n$ is the density of the hydrogen nuclei known from the measured gas pressure and temperature, and $\Delta L$ is the effective target length. The target length was determined with an uncertainty of $\sim$1\%~\cite{Neumaier02}.

\begin{figure}[htbp]
	\centering
	\psfig{file=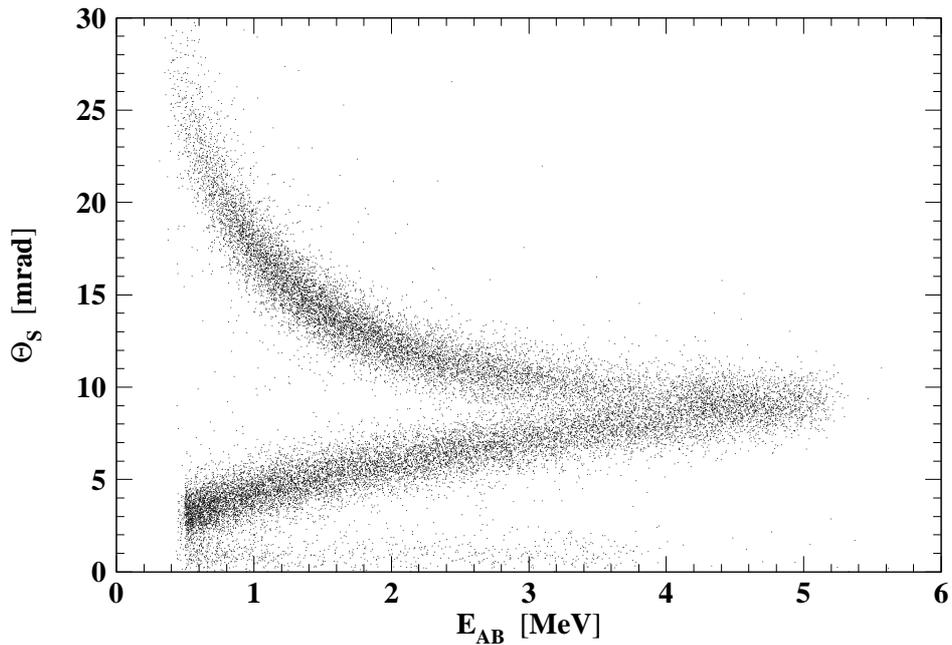,scale=0.7}
	\caption{Correlation between the recoil proton energy $E_{\rm {IKAR}}$ measured in IKAR and the scattering angle $\varTheta_{\rm S}$ of the \8B projectile.}
	\label{IKAR_PC}
\end{figure}

For elastic scattering, the value of $t$ can be obtained in two ways, either from the energy $T_{\rm R}$ of the recoil protons as measured in IKAR, according to
\begin{equation}
- t = 2 m T_{\rm R}~,
\end{equation}
or from the respective projectile scattering angle $\varTheta_{\rm S}$ as determined from the tracking detector system, according to
\begin{equation}
- t = 4 p_{\rm i}^2 \sin^2 \frac{\varTheta_{\rm S}}{2}
\left(1 - \frac{E_{\rm i} T_{\rm R}}{p_{\rm i}^2} \right) \\
\approx  (p_{\rm i}\varTheta_{\rm S})^2 ~~\rm{for ~small }~\varTheta_{\rm S}~.
\end{equation}
Here, $m$, $p_{\rm i}$, and $E_{\rm i}$ denote the proton mass, the projectile initial momentum and the projectile initial total energy, respectively.
Note that the recoil energy \TR~ is connected to the scattering angle $\varTheta_{\rm S}$ by the relation
\begin{equation}
T_{\rm R} = \frac {2 p_{\rm i}^2 \sin^2 \frac{\varTheta_{\rm S}}{2}}
  {m + 2 E_{\rm i} \sin^2 \frac{\varTheta_{\rm S}}{2}}~.
\end{equation}

The major stages in the data analysis, such as the $t$-scale calibration, the determination of the active volume in IKAR, the alignment procedure of the multi-wire proportional chambers, and the selection of the elastic events, were the same as in the previous experiments with IKAR~\cite{Neumaier02,Dobrov06,Ilieva12}. To reject background events, the correlation between the recoil energy $E_{\rm {IKAR}}$ deposited in IKAR and the scattering angle $\varTheta_{\rm S}$ of the projectile was used (Fig.~\ref{IKAR_PC}).
Note that the \8B nucleus has no particle-stable excited states. So in selecting elastic $p^8$B scattering events we had no problem due to a possible contribution of inelastic scattering. As for $p^7$Be scattering, according to our calculations, the contribution of scattering with \7Be excitation in our $t$-range was less than 1\% and could be neglected.
The calculations of the inelastic cross sections were performed using the eikonal approximation and assuming that the nuclear densities have a Gaussian shape, while the transition densities are those from the Tassie model~\cite{Tassie56}. Details of the calculation formalism will be published elsewhere~\cite{Colo18}.

The recoil proton energy \TR~ was calculated in a fairly large $t$-range from the measured scattering angles $\varTheta_{\rm S}$ according to Eq. (4). In the region of small momentum transfers $|t| \lesssim 0.01$ (GeV/$c)^2$, where \TR~$\lesssim 5$ MeV, elastic events correspond to recoil protons stopped within the volume of IKAR. In this case the two ways of calibration of the $t$-scale, according to Eqs. (2) and (3) gave consistent results. In this $t$-region, the recoil energy is measured in IKAR with much higher accuracy than the one determined from the scattering angle $\varTheta_{\rm S}$ of the projectile using Eq. (4)~\cite{Neumaier02}. Therefore, for evaluation of the elastic scattering cross sections \dsdt, the value $t$ as determined from the energy $E_{\rm IKAR}$ measured in IKAR according to Eq. (2) was favoured. In the region of $|t| \geq 0.01$ (GeV/$c$)$^2$ the energy $E_{\rm {IKAR}}$ is only a part of the total recoil energy $T_{\rm R}$, so the determination of the $t$-value from the scattering angle $\varTheta_{\rm S}$ was more accurate and was consequently preferred. The absolute differential cross sections \dsdt~ deduced in the present experiment for proton elastic scattering from the \7Be and \8B isotopes are displayed in Fig.~\ref{crs} and listed in a tabular form in Appendix A as functions of the four-momentum transfer squared $-t$. The indicated energies $E_{\rm p}$ correspond to the equivalent proton energies in direct kinematics: $E_{\rm p} = (E_{\rm i} - M) m / M$, where $M$ is the mass of the projectile. Only statistical errors are given. A high detection efficiency in the present experiment for the beam particles and the elastic-scattering events in the active target IKAR provide the 2\% accuracy of the absolute normalization of the measured cross sections. The uncertainty in the $t$-scale calibration is estimated to be about 1.5\%.

\begin{figure}[h]
	\centering
	\psfig{file=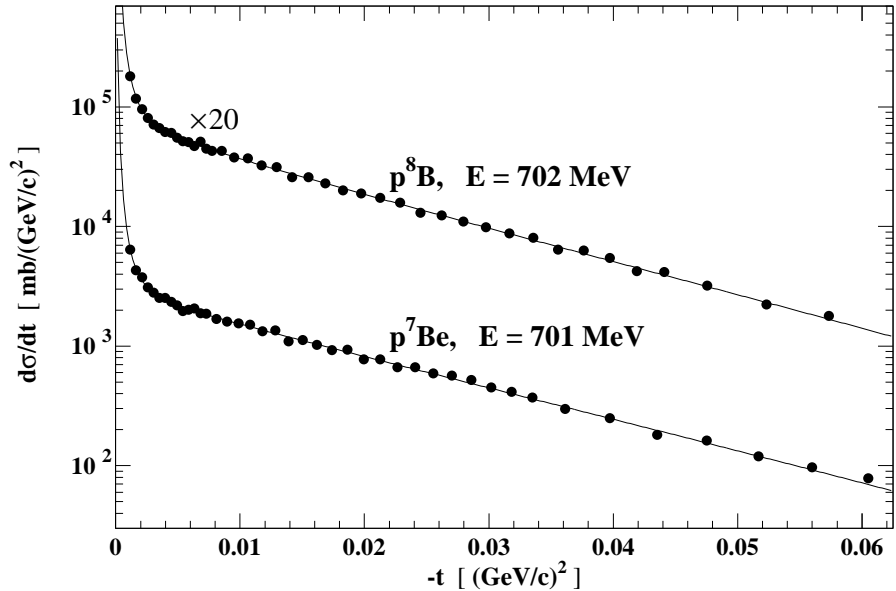,scale=0.66} 
	\caption{Absolute differential cross sections \dsdt~ for $p^7$Be and $p^8$B elastic scattering versus the four-momentum transfer squared. The indicated energies correspond to the equivalent proton energies for direct kinematics. The plotted error bars denote statistical errors only. Solid lines are cross sections calculated within the Glauber theory using the GO parameterization with the fitted parameters.} 
	\label{crs}
\end{figure}


To establish the nuclear density distribution from the measured cross section, the Glauber multiple-scattering theory was applied similarly as in Refs.~\cite{Alk02,Egelhof02,Dobrov06,Ilieva12}. The differential cross sections for proton elastic scattering on composite targets were calculated as
\begin{equation}
{\rm d}\sigma /{\rm d}t = (\pi/k^2)\,|F_{\rm el}(\vq)|^2
\end{equation}
with the scattering amplitude  $F_{\rm el}(\vq)$ given by
\begin{eqnarray}
\quad \quad \quad F_{\rm el}(\vq) & = & ( {\rm i}k/2\pi) \int
{\rm e}^{{\rm i} \vq\vb}\,
\{1 \, - \,\prod^{\rm A}_{\rm i=1}[1-\gamma_{\rm pN}(\vb -
\vs_{\rm i})]\}
\nonumber\\
& \times & \rho_{\rm A}(\vr_1,\vr_2,\ldots,\vr_{\rm A})\,
{\rm d^3}\vr_1~{\rm d^3}\vr_2
\ldots {\rm d^3}\vr_{\rm A}~{\rm d^2}\vb~.
\end{eqnarray}
Here $\vq$ is the momentum transfer ($t = -\vq^2$), $k$ is the wave number of the incident proton, $\vb$ stands for the impact vector, $\gamma_{\rm pN}(\vb)$ represent profile functions for the pairwise $pN$ interactions ($N = p,n$), $A$ is the nuclear mass number, $\vs_{\rm i}$ ($i=1, 2,...,A$) are the transverse nucleon coordinates $\left[ \vr_i\equiv (\vs_i,z_i) \right]$, and $\rho_{\rm A}(\vr_{\rm 1},\vr_{\rm 2},...,\vr_{\rm A})$ denotes the nuclear many-body density. The profile functions $\gamma_{\rm pN}(\vb)$ are related to the corresponding amplitudes $f_{\rm pN}(\vq)$ of free proton-proton $(pp)$ and proton-neutron $(pn)$ scattering by a two-dimensional Fourier transformation. Only the scalar part of the elementary $pN$ scattering amplitude was taken into account, which was described by the high-energy parameterization as
\begin{equation}
f_{\rm pN}(\vq) = ({\rm i}
k/4 \pi)\,\sigma_{\rm pN}~(1\,-\,{\rm i} \varepsilon_{\rm pN})~
{\rm exp}(-\vq^2 \beta_{\rm pN}/2)~,
\end{equation}
where $\sigma_{\rm pp}$ and $\sigma_{\rm pn}$ are the total $pp$ and $pn$ cross sections, $\varepsilon_{\rm pp}$ and $\varepsilon_{\rm pn}$ are the ratios of the real to imaginary parts of the amplitudes, and $\beta_{\rm pp}$ and $\beta_{\rm pn}$ are the slope parameters. The procedure for obtaining the parameters of the free scattering amplitudes is described in detail in Ref.~\cite{Alk02}. The values of these parameters which have been taken as inputs into the present analysis are listed in Table~\ref{params}. In the analysis, the nuclear many-body densities $\rho_{\rm A}$ were taken as products of the one-body densities, which were parameterized with different functions. The parameters of these densities were determined by fitting the calculated cross sections to the experimental data similarly as in the previous experiments~\cite{Alk02,Egelhof02,Dobrov06,Ilieva12}.

\begin{table}[b]
	\centering
	\caption{Values of the free $pp$ and $pn$ amplitudes used in the present analysis of the $p^7$Be and $p^8$B elastic scattering cross sections.
		 $E_{\rm p}$ denotes the equivalent proton energy in direct kinematics.}
	\label{params}
	\vspace*{3pt}
	\begin{tabular}{ccccccc}
		\hline
		\vspace*{3pt}
		Nucleus & $E_{\rm p}$, MeV & $\sigma_{\rm pp}$, mb & $\sigma_{\rm pn}$, mb & $\varepsilon_{\rm pp}$ & $\varepsilon_{\rm pn}$ & $\beta_{\rm pp} = \beta_{\rm pn}$, fm$^2$ \\ \hline
		\vspace*{3pt}
		\7Be   &      701.1       &         43.48         &         37.6          &         0.096          &       $- 0.297$        &                   0.17                    \\
		\8B   &      701.8       &         43.53         &         37.6          &         0.095          &       $- 0.297$        &                   0.17                    \\ \hline
	\end{tabular}
\end{table}

In the present analysis, four parameterizations of phenomenological nuclear density distributions were applied, labeled as SF (Symmetrized Fermi), GH (Gaussian-Halo), GG (Gaussian-Gaussian) and GO (Gaussian-Oscillator). Each of these parameterizations has two free parameters. In the SF parameterization, the free parameters are the \textquotedblleft half density radius\textquotedblright $R_{\rm 0}$ and the diffuseness parameter $a$. The corresponding rms matter radius \Rm \ is given by
\begin{equation}
R_{\rm m} = (3/5)^{1/2}\, R_0~[1 + (7/3)(\pi a/R_0)^2]^{1/2}~.
\end{equation}

\noindent The GH parameterization is determined as a function of the matter radius \Rm \ and the halo parameter $\alpha$, which varies from 0 to 0.4. The case $\alpha$ = 0 corresponds to a Gaussian shape and the one with $\alpha$ = 0.4 to a distribution with a pronounced halo component. While the SF and GH parameterizations do not make any difference between core and halo distributions, the GG and GO parameterizations assume that the nuclei consist of core nucleons and valence nucleons with different spatial distributions. The core distribution is assumed to be a Gaussian one in both the GG and GO parameterizations. The valence nucleon density is described by a Gaussian or a $1p$ shell harmonic oscillator-type distribution within the GG or GO parameterization, respectively. The free parameters in the GG and GO parameterizations are the rms radii $R_{\rm c}$ and $R_{\rm v}$ ($R_{\rm h}$) of the core and valence (\textquotedblleft halo\textquotedblright) nucleon distributions. It was assumed that \8B consists of a \7Be core and a loosely bound valence proton while \7Be was considered to consist of a $^4$He core and a \textquotedblleft halo\textquotedblright composed of two protons and one neutron. The explicit expressions for the SF, GH, GG and GO parameterizations are given in Ref.~\cite{Alk02}.

\section{Results on the nuclear matter density distributions and radii}

\begin{table}[t]
	\centering
	\caption{Summary of the parameters obtained by fitting the calculated $p^7$Be and $p^8$B elastic scattering cross sections to the measured ones for the parameterizations SF, GH, GG and GO of the nuclear matter density distributions. The presented parameters refer to point-nucleon density distributions.
		$A_{\rm n}$ denotes the normalization factor of the calculated cross section (see Ref.~\cite{Alk02}), $N$ is the number of degrees of freedom.
		$A_{\rm n}$, $\chi^2/N$ and $\alpha$ are dimensionless, all other fit parameters are given in fm. The radii $R_{\rm c}$ and $R_{\rm h}$ are in the c.m. system of the nucleus.}
	\label{fit_results}
	\vspace*{2pt}
	\begin{tabular}{cccllll}
		\hline
		\multirow{2}{*}{Nucleus}          & \multirow{2}{*}{Parameterization} & \multicolumn{1}{c}{\multirow{2}{*}{$\chi^2/N$}} &                         \multicolumn{3}{c}{Fit parameters}& \multicolumn{1}{c}{\multirow{2}{*}{\begin{tabular}[c]{@{}c@{}}
					$R_{\rm m}$, \\
					fm
		\end{tabular}}} \\ \cline{4-6}
		&                                  &              \multicolumn{1}{c}{}               & \multicolumn{1}{c}{$A_{\rm n}$} &      \multicolumn{2}{c}{Density parameters}      &  \\ \hline
		\multicolumn{1}{c}{\multirow{4}{*}{\7Be}} &                SF                &                     33.1/44                     & 0.97(1)                         & $R_{\rm 0}$ = 1.23(8) & a = 0.61(2)            & 2.45(3)                             \\
		\multicolumn{1}{l}{}            &                GH                &                     33.1/44                     & 0.97(1)                         & $R_{\rm m}$ = 2.43(3)   & $\alpha$ = 0.08(2)     & 2.43(3)                             \\
		\multicolumn{1}{l}{}            &                GG                &                     33.3/44                     & 0.97(1)                         & $R_{\rm c}$ = 1.95(4)   & $R_{\rm h}$ = 2.94(8)  & 2.42(3)                             \\
		\multicolumn{1}{l}{}            &                GO                &                     33.8/44                     & 0.96(1)                         & $R_{\rm c}$ = 1.76(3)   & $R_{\rm h}$ = 3.06(6)  & 2.40(3)                             \\
		\multicolumn{1}{l}{}            &       \multicolumn{1}{l}{}       &                                                 &                                 &                         &                        &  \\
		\multirow{4}{*}{\8B}            &                SF                &                     26.5/39                     & 0.98(1)                         & $R_{\rm 0}$ = 0.96(31)  & a = 0.66(3)            & 2.57(2)                             \\
		&                GH                &                     25.9/39                     & 0.98(1)                         & $R_{\rm m}$ = 2.56(3)   & $\alpha$ = 0.10(2)     & 2.56(3)                             \\
		&                GG                &                     27.0/39                     & 0.99(1)                         & $R_{\rm c}$ = 2.27(1)   & $R_{\rm h}$ = 4.34(28) & 2.62(6)                             \\
		&                GO                &                     25.9/39                     & 0.98(1)                         & $R_{\rm c}$ = 2.23(1)   & $R_{\rm h}$ = 4.33(20) & 2.59(4)                             \\ \hline
	\end{tabular}
\end{table}

The results of the data analysis with the phenomenological density distributions SF, GH, GG and GO for \7Be and \8B are presented in Table~\ref{fit_results}. For each density parameterization, the deduced rms nuclear matter radii \Rm, the reduced values of $\chi^2$ of the fitting procedure, the values for the fit parameters and the normalization coefficients $A_{\rm n}$ with which the calculated cross sections should be multiplied to obtain the same absolute normalization as the experimental ones are presented. For \7Be and \8B, good descriptions of the cross sections have been achieved with all density parameterizations used. The corresponding values of the rms matter radii \Rm~ deduced with all four parameterizations for \7Be and \8B are close to each other within rather small errors. The values of $R_{\rm m}$ averaged over the results obtained with all density parameterizations are:

\begin{center} 
	$R_{\rm m} = (2.42 \pm 0.04)$ fm  \hspace{1.5cm}for \7Be,

	$R_{\rm m} = (2.58 \pm 0.06)$ fm  \hspace{1.5cm}for \8B,
\end{center}

\noindent where the errors include statistical and systematical uncertainties. The systematical errors in \Rm \ appear due to uncertainties in the absolute normalization of the cross sections, in the $t$-scale calibration, in the parameters of the elementary proton-nucleon scattering amplitudes and due to different model density parameterizations used (for details see Ref.~\cite{Alk02}). The mean value for the core radius of \8B deduced with the GG and GO parameterizations is $R_{\rm c} = 2.25 (3)$ fm. Combining the obtained values of $R_{\rm m}$ and $R_{\rm c}$ and employing the relation between the rms radii \Rm, $R_{\rm c}$, and $R_{\rm h}$
\begin{equation}
{A {R_{\rm m}}^2 = A_{\rm c} {R_{\rm c}}^2 + A_{\rm h} {R_{\rm h}}^2}~,
\end{equation}

\noindent one derives for the halo radius of \8B the value of $R_{\rm h} = 4.24 (25)$ fm.

\begin{figure}[t]
	\centering
	\psfig{file=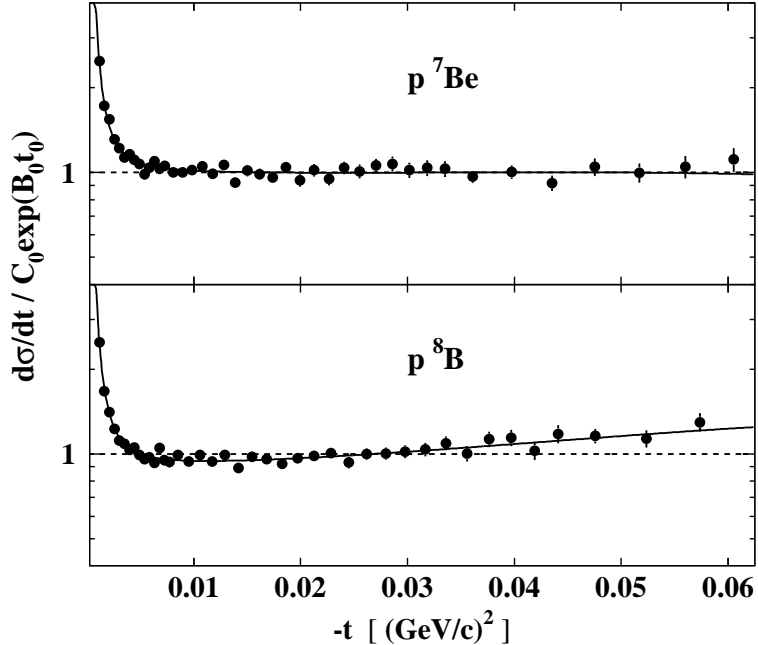,scale=0.7} 
	\caption{The same cross sections as in Fig.~\ref{crs} but divided by an exponential function. The positive curvature in ln(\dsdt) for \8B is a fingerprint for the halo structure of \8B (see text for details).}
	\label{crs2}
\end{figure}

The solid lines in Fig.~\ref{crs} represent the results of the cross sections \dsdt~ using the GO parameterization. At $|t| < 0.005$~(GeV/$c$)$^2$ the steep rise of the cross section with decreasing $|t|$ is caused by Coulomb scattering. The behaviour of the measured curvature of the differential cross section for \8B at $0.005 < |t| < 0.06$~(GeV/$c)^2$ is an indication of the halo occurrence (for details see Refs.~\cite{Alk02,Dobrov06,Korolev18}). This curvature can be better seen if one plots the cross section divided by the exponential function $C_{\rm 0} \cdot \textrm{exp}(B_{\rm 0}t)$, where $B_{\rm 0}$ and $C_{\rm 0}$ are the slope and absolute values of the nuclear part of the differential cross section calculated at $|t| = 0.01$~(GeV/$c$)$^2$.
Such a plot is presented in Fig.~\ref{crs2} for \7Be and \8B using the GO parameterization. As it was shown in
~\cite{Alk02,Dobrov06}, the halo nuclei demonstrate a positive curvature in the
$t$-dependence of ln(\dsdt). This may be explained by the fact that contributions to the cross section for proton scattering from the core and from the halo of these nuclei exhibit different angular dependence. The contribution to the cross section from the scattering on the halo proton decreases faster with increasing $|t|$ than the one from scattering on the core nucleons. In Fig.~\ref{crs2} the fit to the experimental data in the case of \ \8B demonstrates a positive curvature (at $0.01 < |t| < 0.03$ (GeV/$c$)$^2$). This positive curvature in ln(\dsdt) for \8B is an indication of a halo structure. No such positive curvature is observed for \7Be.

\begin{figure}[t]
	\centering
\psfig{file=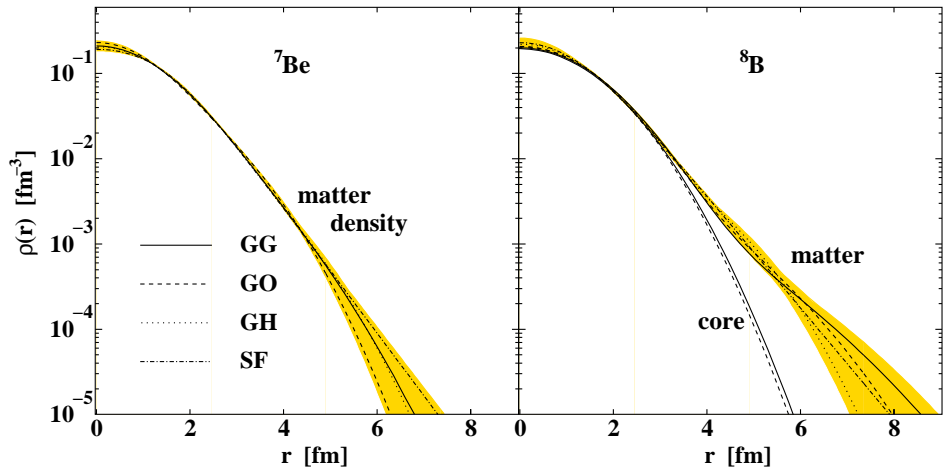,scale=0.45} 
	\caption{Nuclear matter density distributions for \7Be and \8B deduced from the experimental data. In the data analysis, the \8B nucleus was assumed to consist of a \7Be core and a loosely bound valence proton. The error-band represents the envelopes of the density variation within the model parameterizations applied, superimposed by the statistical errors. The results for all parameterizations used in the description of the nuclear matter densities are rather similar and demonstrate clear evidence for the existence of a proton halo in \8B. The deduced core radius in \8B is smaller than the \7Be matter radius in accordance with the theoretical prediction of Grigorenko $et$ $al$.~\cite{Grigorenko98}.} 
	\label{dens}
\end{figure}

The core and nuclear matter distributions deduced for \8B by using different parameterizations of the nuclear matter distributions are compared in Fig.~\ref{dens} with the nuclear matter distribution obtained for \7Be. All density distributions refer to point nucleon distributions. Note that the description of the nuclear matter density distribution for all four parameterizations used are rather similar in the case of \7Be as well as in the case of \8B. The deduced rms matter radii \Rm \ are practically the same for the four versions of the analysis. All versions also resemble each other in reproducing an extended nuclear matter distribution in \8B.

\section{Discussion}

The value of $R_{\rm m} = 2.42 (4)$~fm for \7Be derived in this work coincides within the errors with the result of Warner $et$ $al$.~\cite{Warner01}, but exceeds essentially the value of $R_{\rm m} = 2.31 (2)$~fm obtained by Tanihata $et$ $al$.~\cite{Tanihata88}. A comparison of the rms radii $R_{\rm m}$ and $R_{\rm p}$ calculated using different theoretical models together with the experimental values is presented in Table~\ref{be7_radii}. The $R_{\rm m}$ value from the present experiment agrees within the error limits with most of theoretical results. Note that the best agreement with theoretical calculations is obtained in cases when the experimental value of $R_{\rm p}$~\cite{Lu13,Nortershauser09} measured with a high precision also coincides with the theoretical ones~\cite{Patra98,Carlson15}. The value of $R_{\rm p}$ derived from the interaction cross section measurement~\cite{Tanihata88} is model dependent and is much smaller than that obtained in the laser spectroscopy measurement~\cite{Lu13,Nortershauser09}. By combining the matter radius $R_{\rm m}$, deduced in the present work for \7Be, with $R_{\rm p}$~\cite{Lu13,Nortershauser09} and using the expression
\begin{equation}
{A {R_{\rm m}}^2 = Z {R_{\rm p}}^2 + N {R_{\rm n}}^2}~,
\end{equation}
where $Z$ and $N$ are the numbers of protons and neutrons, the rms neutron radius for \7Be was determined to be
\begin{center} 
	$R_{\rm n} = (2.27 \pm 0.10)$ fm.
\end{center} 

For the thickness of the proton skin $\delta_{\rm pn} = R_{\rm p} - R_{\rm n}$, we deduced the value of $\delta_{\rm pn} = 0.23 (10)$~fm. This result is an indication of a noticeable proton skin in \7Be.

\begin{table}[]
	\centering
	\begin{threeparttable}
		\caption{Comparison of the calculated rms radii of the nuclear matter and the proton densities with the experimental values for $^7$Be. The radii refer to the point-nucleon density distributions.}
		\label{be7_radii}
		\vspace*{4pt}
		\begin{tabular}{cccll}
			\hline
			\multicolumn{1}{l|}{}    & \multicolumn{1}{c}{\Rm, (fm)} & \multicolumn{1}{c|}{\Rp, (fm)} & \multicolumn{2}{c}{Reference}     \\ \hline
			
			\multirow{4}{*}{Experiment}     &       \multicolumn{1}{|c}{\textbf{2.42(4)}}       & --       & \multicolumn{1}{|l}{\textbf{this work}\tnote{a}} &  \\
			& \multicolumn{1}{|c}{2.31 (2)} & 2.36 (2) & \multicolumn{1}{|l}{Tanihata 1988}      & \cite{Tanihata88}\tnote{b} \\
			& \multicolumn{1}{|c}{2.36 (6)} &  --      & \multicolumn{1}{|l}{Warner 2001}        & \cite{Warner01}\tnote{c}   \\
			& \multicolumn{1}{|c}{ --    }  &2.507 (17)& \multicolumn{1}{|l}{N\"{o}rtersh\"{a}user 2009}    & \cite{Nortershauser09}\tnote{d}  \\
			\hline
			\multirow{12}{*}{Theory}    &  \multicolumn{1}{|c}{2.43}  &  2.48  &  \multicolumn{1}{|l}{Cs\'{o}t\'{o} 1993}      & \cite{Csoto93}\tnote{e} \\
			&  \multicolumn{1}{|c}{2.36}  &  2.41  &  \multicolumn{1}{|l}{Varga 1995}      & \cite{Varga95}\tnote{e}       \\
			&  \multicolumn{1}{|c}{2.420} &  2.549 &  \multicolumn{1}{|l}{Fayans 1995}     & \cite{Fayans95}\tnote{f}      \\
			&  \multicolumn{1}{|c}{2.50}  &  2.64  &  \multicolumn{1}{|l}{Shen 1996}       & \cite{Shen96}\tnote{g}        \\
			&  \multicolumn{1}{|c}{2.280} &  2.369 &  \multicolumn{1}{|l}{Negoita 1996}    & \cite{Negoita96}\tnote{g}     \\
			&  \multicolumn{1}{|c}{2.37--2.40}& --  &  \multicolumn{1}{|l}{Grigorenko 1998} & \cite{Grigorenko98}\tnote{e}  \\
			&  \multicolumn{1}{|c}{2.413} &  2.525 &  \multicolumn{1}{|l}{Patra 1998}      & \cite{Patra98}\tnote{h}       \\
			&  \multicolumn{1}{|c}{2.49}  &  2.63  &  \multicolumn{1}{|l}{Dhiman 2005}     & \cite{Chandel03}\tnote{g}     \\
			&  \multicolumn{1}{|c}{2.327} &  2.455 &  \multicolumn{1}{|l}{Wang 2009}       & \cite{Wang09}\tnote{h}        \\
			&  \multicolumn{1}{|c}{2.37}  &  2.45  &  \multicolumn{1}{|l}{Krieger 2012}    & \cite{Krieger12}\tnote{i}     \\
			&  \multicolumn{1}{|c}{2.39}  &  2.47  &  \multicolumn{1}{|l}{Pastore 2013}    & \cite{Pastore13}\tnote{j}     \\
			&  \multicolumn{1}{|c}{2.43}  &  2.51  &  \multicolumn{1}{|l}{Carlson 2015}    & \cite{Carlson15}\tnote{j}     \\
			\hline
		\end{tabular}
		\begin{tablenotes}
			\item[a] proton elastic scattering
			\item[b] interaction cross section measurements
			\item[c] reaction cross section measurements
			\item[d] laser spectroscopy measurements
			\item[e] microscopic cluster model
			\item[f] semi-microscopic folding calculations
			\item[g] Skirme-Hartree-Fock (SkHF) model
			\item[h] Relativistic Mean Field (RMF) model
			\item[i] Fermionic Molecular Dynamics (FMD) calculations
			\item[j] Green's function Monte Carlo (GFMC) calculations
		\end{tablenotes}
	\end{threeparttable}
\end{table}

A summary of the results on the structure of the proton-rich \8B nucleus obtained from experimental and theoretical research is presented in Table~\ref{b8_radii}. The experimental value of $R_{\rm m} = 2.58 (6)$~fm, deduced in the present work, is in good agreement with the value of $R_{\rm m} = 2.61 (8)$ fm obtained with the modified Glauber model approach in the recent analysis~\cite{Fan15} of all existing data on reaction cross sections, but is larger than that of earlier results of Refs. \cite{Tanihata88} and \cite{Obuti96}. The present value of $R_{\rm m}$ turns out to be within the experimental errors in almost perfect agreement with many theoretical results~\cite{Csoto93,Varga95,Grigorenko98,Chandel03} presented in Table~\ref{b8_radii}. In some theoretical studies the nucleon structure of \8B was treated as a three-cluster system~\cite{Csoto93,Baye94,Varga95,Grigorenko98}. In particular, the theoretical description of \8B in Ref.~\cite{Grigorenko98} is performed assuming a ($^4$He + $^3$He + $p$) three-body model with explicit inclusion of the binary ($^7$Be + $p$) channel. The model describes the bulk properties of \8B well and predicts a value of $R_{\rm m}$ = 2.59 fm. The model also reproduces the experimental data~\cite{Smedberg99} on the shape and a rather narrow width of the momentum distribution of the \7Be fragments in the proton removal channel of the \8B high-energy break-up. An important finding of this model~\cite{Grigorenko98} is that the presence of a loosely bound proton leads to a contraction of the \7Be cluster inside \8B. Indeed, according to the present measurements, the deduced \7Be core of the \8B nucleus $R_{\rm c}$ = 2.25(3) fm is essentially smaller than the \7Be matter radius $R_{\rm m}$ = 2.42(4) fm.

\begin{table}[]
	\centering
	\begin{threeparttable}
		\caption{Summary of the results obtained for $^8$B from experimental and theoretical studies. The values \Rm, $R_{\rm p}$ and $R_{\rm n}$ denote the rms radii of the nuclear matter, and of the proton and neutron distributions, respectively, and refer to the point-nucleon density distributions.}
		\label{b8_radii}
		\vspace*{4pt}
		\begin{tabular}{ccccll}
			\hline
			\multicolumn{1}{l|}{}    & \multicolumn{1}{c}{\Rm, (fm)} & \multicolumn{1}{c}{\Rp, (fm)} & \multicolumn{1}{c|}{$R_{\rm n}$, (fm)} &     \multicolumn{2}{c}{Reference}     \\ \hline
			
			\multirow{7}{*}{Experiment} &      \multicolumn{1}{|c}{\textbf{2.58 (6)}}       &       \textbf{2.76 (9)}       &           \textbf{2.25 (3)}            & \multicolumn{1}{|l}{\textbf{this work}\tnote{a}} &  \\
			& \multicolumn{1}{|c}{2.38 (4)} & 2.45 (5) & 2.27 (4) & \multicolumn{1}{|l}{Tanihata 1988}      & \cite{Tanihata88}\tnote{b} \\
			& \multicolumn{1}{|c}{2.43 (3)} & 2.49 (3) & 2.33 (3) & \multicolumn{1}{|l}{Obuti 1996}         & \cite{Obuti96}\tnote{b}    \\
			& \multicolumn{1}{|c}{2.50 (4)} &  --      &  --      & \multicolumn{1}{|l}{Al-Khalili 1996}    & \cite{AlKhali96}\tnote{c}  \\
			& \multicolumn{1}{|c}{2.55 (8)} & 2.76 (8) & 2.16 (3) & \multicolumn{1}{|l}{Negoita 1996}       & \cite{Negoita96}\tnote{d}  \\
			& \multicolumn{1}{|c}{2.45(10)} & 2.53 (13)& 2.31 (5) & \multicolumn{1}{|l}{Fukuda 1999}        & \cite{Fukuda99}\tnote{c}   \\
			& \multicolumn{1}{|c}{2.61 (8)} &  --      &  --      & \multicolumn{1}{|l}{Fan 2015}           & \cite{Fan15}\tnote{e}      \\
			\hline
			\multirow{17}{*}{Theory}    &  \multicolumn{1}{|c}{2.71}  &  2.98  &  2.20   & \multicolumn{1}{|l}{Minamisono 1992}      & \cite{Minamisono92}\tnote{f} \\
			&  \multicolumn{1}{|c}{2.57}  &  2.74  &  2.25   & \multicolumn{1}{|l}{Cs\'{o}t\'{o} 1993}   & \cite{Csoto93}\tnote{g}      \\
			&  \multicolumn{1}{|c}{2.73}  &  2.88  &  2.46   & \multicolumn{1}{|l}{Baye 1994}            & \cite{Baye94}\tnote{g}       \\
			&  \multicolumn{1}{|c}{2.56}  &  2.73  &  2.24   & \multicolumn{1}{|l}{Varga 1995}           & \cite{Varga95}\tnote{g}      \\
			&  \multicolumn{1}{|c}{2.507} &  2.68  &  2.19   & \multicolumn{1}{|l}{Fayans 1995}          & \cite{Fayans95}\tnote{h}     \\
			&  \multicolumn{1}{|c}{2.68}  &  2.92  &  2.21   & \multicolumn{1}{|l}{Brown 1996}           & \cite{Brown96}\tnote{f}      \\
			&  \multicolumn{1}{|c}{2.59}  &  2.75  &  2.30   & \multicolumn{1}{|l}{Grigorenko 1998}      & \cite{Grigorenko98}\tnote{g} \\
			&  \multicolumn{1}{|c}{2.494} &  2.654 &  2.202  & \multicolumn{1}{|l}{Patra 1998}           & \cite{Patra98}\tnote{j}      \\
			&  \multicolumn{1}{|c}{2.627} &  2.861 &  2.181  & \multicolumn{1}{|l}{Kitagawa 1999}        & \cite{Kitagawa99}\tnote{f}   \\
			&  \multicolumn{1}{|c}{2.57}  &  2.73  &  2.27   & \multicolumn{1}{|l}{Dhiman 2005}          & \cite{Chandel03}\tnote{i}    \\
			&  \multicolumn{1}{|c}{2.50}  &  2.64  &  2.24   & \multicolumn{1}{|l}{Furutachi 2009}       & \cite{Furutachi09}\tnote{k}  \\
			&  \multicolumn{1}{|c}{2.367} &  2.537 &  2.052  & \multicolumn{1}{|l}{Wang 2009}            & \cite{Wang09}\tnote{j}       \\
			&  \multicolumn{1}{|c}{2.35}  &  2.48  &  2.11   & \multicolumn{1}{|l}{Pastore 2013}         & \cite{Pastore13}\tnote{l}    \\
			&  \multicolumn{1}{|c}{2.262} &  2.373 &  2.062  & \multicolumn{1}{|l}{Henninger 2015}       & \cite{Henninger15}\tnote{m}  \\
			\hline
		\end{tabular}
		\begin{tablenotes}
			\item[a] proton elastic scattering; $R_{\rm p}$ and $R_{\rm n}$ are the estimations of the rms radii of the proton and neutron distributions in \8B performed in the present work under the assumption that $R_{\rm n}$ = $R_{\rm c}$
			\item[b] Glauber-model analysis of the interaction cross section
			\item[c] Glauber-model analysis of the reaction cross section
			\item[d] analysis of the break-up and reaction cross sections
			\item[e] modified Glauber model used in the analysis of existing reaction cross sections
			\item[f] shell model
			\item[g] microscopic cluster model
			\item[h] semi-microscopic folding calculations
			\item[i] Skirme-Hartree-Fock (SkHF) model
			\item[j] Relativistic Mean Field (RMF) model
			\item[k] an extended framework based on antisymmetrized molecular dynamics (MAMD)
			\item[l] Green's function Monte Carlo (GFMC) calculations
			\item[m] Fermionic Molecular Dynamics (FMD) calculations
		\end{tablenotes}
	\end{threeparttable}
\end{table}

The rms halo radius $R_{\rm h}$ = 4.24(25) fm derived in the present experiment confirms the halo nature of \8B and can be compared with the values obtained in other works (Table~\ref{halo_radii}). Experimental results presented there include measurements of the nuclear asymptotic normalization coefficient (ANC) which permits to extract the halo radius~\cite{Carstoiu01,Liu04,Carstoiu07} and the model-dependent analysis of the break-up cross section~\cite{Negoita96}.
Several theoretical calculations~\cite{Kelley96,Brown96,Grigorenko98,Wang09} are also presented in Table~\ref{halo_radii}. The present value of the halo radius $R_{\rm h}$ agrees within the error limits with the results of all investigations displayed in Table~\ref{halo_radii}.

A criterion for a quantitative assessment of halo nuclei was proposed in Ref.~\cite{Grigorenko98}. The ratio of the valence nucleon to the core radii
$\kappa = R_{\rm v} / R_{\rm c}$ is used as a gauge for the halo existence. For light nuclei close to the valley of beta stability, theory predicts typically
$\kappa \approx 1.20 -1.25$, while for the halo structure this value can be essentially larger, up to $\kappa > 2$~\cite{Jonson04}. In the three-cluster model~\cite{Grigorenko98} a value of $\kappa = 1.75$ was obtained for \8B which can be compared with $\kappa = 1.88 (14)$ deduced from the present measurement. In previous experiments on proton elastic scattering with the same method~\cite{Alk02,Ilieva12}, we obtained for the neutron halo nuclei $^6$He and $^{14}$Be the values of $\kappa = 1.76$ and $\kappa = 1.91$,  respectively, the values which are close to that for \8B.

Up to the present, there are no direct measurements of the nuclear charge radius $R_{\rm ch}$ (or proton radius $R_{\rm p}$) of \8B. An estimation of the rms radius of the proton distribution in \8B can be performed using the results of the present experiment. Assuming that for \8B the rms radius of the neutron distribution $R_{\rm n}$ equals the core radius $R_{\rm c}$ and using expression (10) we have obtained the rms proton radius to be 

\begin{center} 
	$R_{\rm p} = (2.76 \pm 0.09)$ fm.
\end{center}

\begin{figure}[t]
	\centering
	\psfig{file=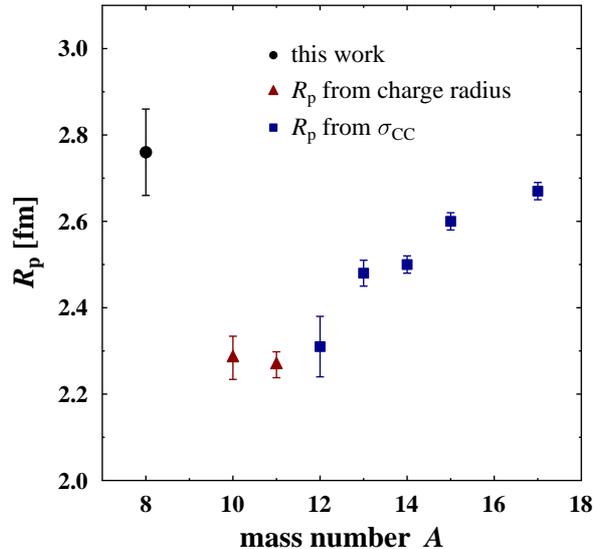,scale=0.60} 
	\caption{Experimental rms proton radii $R_{\rm p}$ for the boron isotopes. The filled circle is the result of the present work. The filled triangles are the values of $R_{\rm p}$ for $^{10}$B and $^{11}$B determined from the measured charge radii~\cite{Angeli13}. The squares are the proton radii $R_{\rm p}$ deduced from the measured charge-changing cross sections using a finite-range Glauber model analysis~\cite{Estrade14}. The radii refer to the point-nucleon density distributions.}
	\label{radii}
\end{figure}
\begin{table}[b]
	\centering
	\caption{Experimental and theoretical values of the rms halo radius $R_{\rm h}$ in $^8$B.}
	\label{halo_radii}
	\vspace*{4pt}
	\begin{tabular}{llll}
		\hline
		\multicolumn{1}{c}{$R_{\rm h}$, (fm)} & \multicolumn{2}{c}{Reference} & \multicolumn{1}{c}{}     \\ \hline
		
		\multirow{2}{*}{4.62 (24)} & \multirow{2}{*}{Carstoiu 2007} & \multirow{2}{*}{\cite{Carstoiu07}} & {measurement of the asymptotic} \\
		{ }         &                      &                    & normalization coefficient (ANC) \\
		{4.44 }     & {Brown 1996}         & \cite{Brown96}     & shell-model  calculations \\
		{4.24 }     & {Kelley 1996}        & \cite{Kelley96}    & single-particle Hamiltonian model \\
		\textbf{4.24 (25)}  & \textbf{This work}   &                    & proton elastic scattering       \\
		{4.20 (22)} & {Carstoiu 2001}      & \cite{Carstoiu01}  & measurement of the ANC     \\
		{4.03}      & {Grigorenko 1998}    & \cite{Grigorenko98}& cluster model       \\
		{3.98}      & {Wang 2009}          & \cite{Wang09}      & ANC method in RMF theory \\
		{3.97 (12)} & {Negoita 1996}       & \cite{Negoita96}   & model-dependent break-up cross section analysis \\
		{3.90 (20)} & {Liu 2004}           & \cite{Liu04}       & measurement of the ANC \\
		\hline
	\end{tabular}
\end{table}

\noindent This value may be underestimated because we have admitted that the proton and neutron rms radii of the core in \8B are the same, whereas the core proton radius is expected to be slightly larger than the neutron one, as it is the case in the free \7Be nucleus, and for this reason the deduced $R_{\rm p}$ value may be larger by $\sim$0.05 fm. The results of the present estimation on $R_{\rm p}$ (and $R_{\rm n}$) for \8B are compared in Table~\ref{b8_radii} with results of other experimental and theoretical investigations. The values indicated as experimental are model dependent and are widely scattered. The present estimation of $R_{\rm p}$ is in agreement with some of the theoretical calculations~\cite{Csoto93,Varga95,Grigorenko98,Chandel03}. In Fig.~\ref{radii}, the present estimation of the $R_{\rm p}$ value for \8B is compared with the existing experimental data on $R_{\rm p}$ values for other boron isotopes. The weighted average charge radii $R_{\rm ch} = 2.43 (5)$~fm and $R_{\rm ch} = 2.41 (3)$~fm for the stable $^{10}$B and $^{11}$B are known from electron and $\pi^+$ scattering measurements and from muonic atom X-rays studies~\cite{Angeli13}. Taking into account the relation between the point proton and the charge radius of a nucleus~\cite{Lu13}, the corresponding proton radii are deduced to be $R_{\rm p} = 2.28 (5)$~fm for $^{10}$B, and $R_{\rm p}$ = 2.27(3)~fm for $^{11}$B, respectively. The values $R_{\rm p}$ for the neutron-rich boron isotopes $^{12-17}$B were recently determined from charge-changing cross section measurements at GSI~\cite{Estrade14}.
The value of $R_{\rm p}$ for \8B determined in the present work is fairly larger than those of nuclei of the stable isotopes $^{10}$B and $^{11}$B, which confirms the existence of a halo in \8B. Using the estimated value of $R_{\rm p}$ we have also deduced the charge radius for \8B as $R_{\rm ch} = 2.89 (9)$~fm.

\section{Summary}

In the present work we used a method, developed by PNPI--GSI collaboration, of small angle proton-nucleus elastic scattering in inverse kinematics to determine the nuclear matter density distribution of the proton-rich \7Be and \8B nuclei. The absolute differential cross sections for proton elastic scattering was measured in the ranges $0.001 \leqslant |t| \leqslant 0.08$~(GeV/$c$)$^2$ and $0.001 \leqslant |t| \leqslant 0.06$ (GeV/$c)^2$ of the four-momentum transfer squared for $p^7$Be, and $p^8$B scattering, respectively. The cross sections were determined using secondary beams from the GSI fragment separator FRS at an energy of $\sim700$ MeV/nucleon. The hydrogen-filled ionization chamber IKAR served simultaneously as a hydrogen target and a recoil-proton detector. The nuclear matter radii and radial matter distributions were determined with the aid of the Glauber multiple-scattering theory. In the analysis, four phenomenological parameterizations of the nuclear density distributions were used, each of these parameterizations had two free parameters. For both investigated nuclei, a good description of the cross sections was obtained with all the used density parameterizations. The value of $R_{\rm m} = 2.42 (4)$~fm deduced for \7Be is larger than the matter radii obtained in previous experiments based on measurements of the total interaction~\cite{Tanihata88} and reaction~\cite{Warner01} cross sections. On the other hand, the present $R_{\rm m}$ value is in perfect agreement with most of the theoretical predictions (Table~\ref{be7_radii}). A noticeable  proton skin $\delta_{\rm pn} = 0.23 (10)$~fm was obtained combining the value $R_{\rm m}$ with the proton radius $R_{\rm p}$ measured by the laser spectroscopy technique~\cite{Lu13,Nortershauser09}. Similar matter density distributions were determined using the all four parameterizations.

The nuclear matter density distributions deduced for the proton-rich \8B nucleus are also very similar for all the parameterizations used and demonstrate an extended distribution. In the case of the GG and GO parameterizations, the deduced rms halo radius $R_{\rm h} = 4.24 (25)$~fm is 1.88 times larger than the rms core radius $R_{\rm c} = 2.25 (3)$~fm, thus giving clear evidence of a halo structure. The value of $R_{\rm h}$ is in good agreement with the existing experimental measurements and theoretical calculations (Table~\ref{halo_radii}). The nuclear matter radius $R_{\rm m} = 2.58 (6)$~fm deduced in this work is in perfect agreement with the recent experimental result of Fan $et$ $al$.~\cite{Fan15}, where $R_{\rm m}$ was obtained from the existing data on reaction cross sections using a modified Glauber model, and in agreement with several theoretical calculations~\cite{Csoto93,Varga95,Grigorenko98,Chandel03} (Table~\ref{b8_radii}). In the three-body model, Grigorenko $et$ $al$.~\cite{Grigorenko98} predicted a contraction of the \7Be cluster inside \8B. The present measurement supports this finding. Assuming that the rms radius of the neutron distribution $R_{\rm n}$ for \8B is equal to the core radius $R_{\rm c}$, we have performed an estimation of the rms radius of the proton distribution as $R_{\rm p} = 2.76 (9)$~fm. This value is in agreement with theoretical calculations (the same as in the case of $R_{\rm m}$) and is significantly larger than the $R_{\rm p}$ values for the neighbouring successive boron isotopes (Fig.~\ref{radii}).

\section*{Acknowledgements}

The authors are grateful to A.~Bleile, G.~Ickert, A.~Br\"{u}nle, K.-H.~Behr and W.~Niebur for their technical assistance in the preparation of the experimental set-up. The visiting group from PNPI thanks the GSI authorities for the hospitality.

\section*{References}

\bibliography{bibfile}

\begin{thebibliography}{10}
\expandafter\ifx\csname url\endcsname\relax
  \def\url#1{\texttt{#1}}\fi
\expandafter\ifx\csname urlprefix\endcsname\relax\def\urlprefix{URL }\fi
\expandafter\ifx\csname href\endcsname\relax
  \def\href#1#2{#2} \def\path#1{#1}\fi

\bibitem{Tanihata13}
I.~Tanihata, H.~Savajols, R.~Kanungo, Recent experimental progress in nuclear
  halo structure studies, Prog. Part. Nucl. Phys. 68 (2013) 215.

\bibitem{Jonson04}
B.~Jonson, Light dripline nuclei, Phys. Rep. 389 (2004) 1.

\bibitem{Ozawa01}
A.~Ozawa, T.~Suzuki, I.~Tanihata, Nuclear size and related topics, Nucl. Phys.
  A 693 (2001) 32.

\bibitem{Tanihata96}
I.~Tanihata, Neutron halo nuclei, J. Phys. G 22 (1996) 157.

\bibitem{Wang12}
M.~Wang, G.~Audi, A.~Wapstra, F.~Kondev, M.~MacCormick, X.~Xu, B.~Pfeiffer,
  {The Ame2012 atomic mass evaluation}, Chin. Phys. C 36 (2012) 1603.

\bibitem{Minamisono92}
T.~Minamisono, T.~Ohtsubo, I.~Minami, S.~Fukuda, A.~Kitagawa, M.~Fukuda,
  K.~Matsuta, Y.~Nojiri, S.~Takeda, H.~Sagawa, H.~Kitagawa, Proton halo of
  {$^8$B} disclosed by its giant quadrupole moment, Phys. Rev. Lett. 69 (1992)
  2058.

\bibitem{Csoto93}
A.~Cs\'{o}t\'{o}, Proton skin of {$^8$B} in a microscopic model, Phys. Lett. B
  315 (1993) 24.

\bibitem{Nakada94}
H.~Nakada, T.~Otsuka, {$E2$} properties of nuclei far from stability and the
  proton-halo problem of {$^8$B}, Phys. Rev. C 49 (1994) 886.

\bibitem{Schwab95}
W.~Schwab, H.~Geissel, H.~Lenske, K.-H. Behr, A.~Br{\"{u}}nle, K.~Burkard,
  H.~Irnich, T.~Kobayashi, G.~Kraus, A.~Magel, G.~M{\"{u}}nzenberg, F.~Nickel,
  K.~Riisager, C.~Scheidenberger, B.~Sherrill, T.~Suzuki, B.~Voss, Observation
  of a proton halo in {$^8$B}, Z. Phys. A 350 (1995) 283.

\bibitem{Kelley96}
J.~Kelley, S.~Austin, A.~Azhari, D.~Bazin, J.~Brown, H.~Esbensen, M.~Fauerbach,
  M.~Hellstr{\"{o}}m, S.~Hirzebruch, R.~Kryger, D.~Morrissey, R.~Pfaff,
  C.~Powell, E.~Ramakrishnan, B.~Sherrill, M.~Steiner, T.~Suomij{\"{a}}rvi,
  M.~Thoennessen, {Study of the Breakup Reaction {$^8$B} $\rightarrow$ {$^7$Be}
  + $p$: Absorption Effects and $E2$ Strength}, Phys. Rev. Lett. 77 (1996)
  5020.

\bibitem{Negoita96}
F.~Negoita, C.~Borcea, F.~Carstoiu, M.~Lewitowicz, M.~Saint-Laurent, R.~Anne,
  D.~Bazin, J.~Corre, P.~Roussel-Chomaz, V.~Borrel, D.~Guillemaud-Mueller,
  H.~Keller, A.~Mueller, F.~Pougheon, O.~Sorlin, S.~Lukyanov,
  Y.~Penionzhkevich, A.~Fomichev, N.~Skobelev, O.~Tarasov, Z.~Dlouhy,
  A.~Kordyasz, {$^8$B} proton halo via reaction and breakup cross section
  measurements, Phys. Rev. C 54 (1996) 1787.

\bibitem{Smedberg99}
M.~Smedberg, T.~Baumann, T.~Aumann, L.~Axelsson, U.~Bergmann, M.~Borge,
  D.~Cortina-Gil, L.~Fraile, H.~Geissel, L.~Grigorenko, M.~Hellstr{\"{o}}m,
  M.~Ivanov, N.~Iwasa, R.~Janik, B.~Jonson, H.~Lenske, K.~Markenroth,
  G.~M{\"{u}}nzenberg, T.~Nilsson, A.~Richter, K.~Riisager, C.~Scheidenberger,
  G.~Schrieder, W.~Schwab, H.~Simon, B.~Sitar, P.~Strmen, K.~S{\"{u}}mmerer,
  M.~Winkler, M.~Zhukov, New results on the halo structure of {$^8$B}, Phys.
  Lett. B 452 (1999) 1.

\bibitem{Cortina03}
D.~Cortina-Gil, J.~Fernandez-Vazquez, F.~Attallah, T.~Baumann, J.~Benlliure,
  M.~Borge, L.~Culkov, C.~Forss{\'{e}}n, L.~Fraile, H.~Geissel, J.~Gerl,
  K.~Itahashi, R.~Janik, B.~Jonson, S.~Karlsson, H.~Lenske, S.~Mandal,
  K.~Markenroth, M.~Meister, M.~Mocko, G.~M{\"{u}}nzenberg, T.~Ohtsubo,
  A.~Ozawa, Y.~Parfenova, V.~Pribora, A.~Richter, K.~Riisager, R.~Schneider,
  H.~Scheit, G.~Schrieder, N.~Shulgina, H.~Simon, B.~Sitar, A.~Stolz,
  P.~Strmen, K.~S{\"{u}}mmerer, I.~Szarka, S.~Wan, H.~Weick, M.~Zhukov,
  {Nuclear and Coulomb breakup of $^8$B}, Nucl. Phys. A 720 (2003) 3.

\bibitem{Blank97}
B.~Blank, C.~Marchand, M.~Primakoff, T.~Baumann, F.~Bou{\'{e}}, H.~Geissel,
  M.~Hellstr{\"{o}}m, N.~Iwasa, W.~Schwab, K.~S{\"{u}}mmerer, M.~Gai, Total
  interaction and proton-removal cross-section measurements for the proton-rich
  isotopes {$^7$Be}, {$^8$B}, and {$^9$C}, Nucl. Phys. A 624 (1997) 242.

\bibitem{Fukuda99}
M.~Fukuda, M.~Mihara, T.~Fukao, S.~Fukuda, M.~Ishihara, S.~Ito, T.~Kobayashi,
  K.~Matsuta, T.~Minamisono, S.~Momota, T.~Nakamura, Y.~Nojiri, Y.~Ogawa,
  T.~Ohtsubo, T.~Onishi, A.~Ozawa, T.~Suzuki, M.~Tanigaki, I.~Tanihata,
  K.~Yoshida, Density distribution of {$^8$B} studied via reaction cross
  sections, Nucl. Phys. A 656 (1999) 209.

\bibitem{Warner04}
R.~Warner, F.~Becchetti, J.~Brown, A.~Galonsky, J.~Kelley, A.~Nadsen,
  R.~Ronningen, J.~Tostevin, J.~Winfield, P.~Zecher, {Proton removal from
  $^8$B, $^9$C, and $^{12}$C on Si at 20--70 MeV/nucleon}, Phys. Rev. C 69
  (2004) 024612.

\bibitem{Tanihata88}
I.~Tanihata, T.~Kobayashi, O.~Yamakawa, S.~Shimoura, K.~Ekuni, K.~Sugimoto,
  N.~Takahashi, T.~Shimoda, H.~Sato, {Measurement of interaction cross sections
  using isotope beams of Be and B and isospin dependence of the nuclear radii},
  Phys. Lett. B 206 (1988) 592.

\bibitem{Warner95}
R.~Warner, J.~Kelley, P.~Zecher, F.~Becchetti, J.~Brown, C.~Carpenter,
  A.~Galonsky, B.~Sherrill, J.~Wang, J.~Winfield, {Evidence for a proton halo
  in $^8$B: Enhanced total reaction cross sections at 20 to 60 MeV/nucleon},
  Phys. Rev. C 52 (1995) R1166.

\bibitem{AlKhali96}
J.~Al-Khalili, J.~Tostevin, {Matter Radii of Light Halo Nuclei}, Phys. Rev.
  Lett. 76 (1996) 3903.

\bibitem{Obuti96}
M.~Obuti, T.~Kobayashi, D.~Hirata, Y.~Ogawa, A.~Ozawa, K.~Sugimoto,
  I.~Tanihata, D.~Olson, W.~Christie, H.~Wieman, {Interaction cross section and
  interaction radius of the $^8$B nucleus}, Nucl. Phys. A 609 (1996) 74.

\bibitem{Fan15}
G.~Fan, M.~Fukuda, D.~Nishimura, X.~Cai, S.~Fukuda, I.~Hachiuma, C.~Ichikawa,
  T.~Izumikawa, M.~Kanazawa, A.~Kitagawa, T.~Kuboki, M.~Lantz, M.~Mihara,
  M.~Nagashima, K.~Namihira, Y.~Ohkuma, T.~Ohtsubo, Z.~Ren, S.~Sato, Z.~Sheng,
  M.~Sugiyama, S.~Suzuki, T.~Suzuki, M.~Takechi, T.~Yamaguchi, W.~Xu, Density
  distribution of {$^8$Li} and {$^8$B} and capture reaction at low energy,
  Phys. Rev. C 91 (2015) 014614.

\bibitem{Baye94}
D.~Baye, P.~Descouvemont, N.~Timofeyuk, Matter densities of {$^8$B} and
  {$^8$Li} in a microscopic cluster model and the proton-halo problem of
  {$^8$B}, Nucl. Phys. A 577 (1994) 624.

\bibitem{Brown96}
B.~Brown, A.~Cs\'{o}t\'{o}, R.~Sherr, {Coulomb displacement energy and the
  low-energy astrophysical $S_{\rm 17}$ factor for the $^7$Be(p,$\gamma$)$^8$B
  reaction}, Nucl. Phys. A 597 (1996) 66.

\bibitem{Kitagawa99}
H.~Kitagawa, {Shell Model Study of the Quadrupole Moments in Light Mirror
  Nuclei}, Progr. Theor. Phys. 102 (1999) 1015.

\bibitem{Furutachi09}
N.~Furutachi, M.~Kimura, A.~Dot{\'{e}}, Y.~Kanada-En'yo, {Structures of Light
  Halo Nuclei}, Progr. Theor. Phys. 122 (2009) 865.

\bibitem{Henninger15}
K.~Henninger, T.~Neff, H.~Feldmeier, {$^8$B structure in Fermionic Molecular
  Dynamics}, J. Phys.: Conf. Ser. 599 (2015) 012038.

\bibitem{Varga95}
K.~Varga, Y.~Suzuki, I.~Tanihata, Microscopic four-cluster description of the
  mirror nuclei {$^9$Li} and {$^9$C}, Phys. Rev. C 52 (1995) 3013.

\bibitem{Fayans95}
S.~Fayans, O.~Knyazkov, I.~Kuchtina, Y.~Penionzhkevich, N.~Skobelev,
  Quasielastic scattering of light exotic nuclei. a semi-microscopic folding
  analysis, Phys. Lett. B 357 (1995) 509.

\bibitem{Grigorenko98}
L.~Grigorenko, B.~Danilin, V.~Efros, N.~Shu{\v{l}}gina, M.~Zhukov, {Structure
  of the $^8$Li and $^8$B nuclei in an extended three-body model and
  astrophysical $S_{\rm 17}$ factor}, Phys. Rev. C 57 (1998) 2099(R).

\bibitem{Patra98}
S.~Patra, C.-L. Wu, C.~Praharaj, Proton-skin in {$^8$B}-nucleus, Mod. Phys.
  Lett. A 13 (1998) 2743.

\bibitem{Chandel03}
S.~Dhiman, R.~Shyam, {Structure of $^8$B and astrophysical $S_{\rm 17}$
  factor}, J. Phys. G: Nucl. Part. Phys. 31 (2005) S1531.

\bibitem{Wang09}
C.~Wang, T.~Dong, Z.~Zhu, Z.~Ren, {Asymptotic normalization coefficients for
  $^7$Be(p,$\gamma$)$^8$B from RMF calculation}, Mod. Phys. Lett. A 24 (2009)
  1453.

\bibitem{Pastore13}
S.~Pastore, S.~Pieper, R.~Schiavilla, R.~Wiringa, {Quantum Monte Carlo
  calculations of electromagnetic moments and transitions in $A \leq 9$ nuclei
  with meson-exchange currents derived from chiral effective field theory},
  Phys. Rev. C 87 (2013) 035503.

\bibitem{Warner01}
R.~Warner, M.~McKinnon, J.~Needleman, N.~Shaner, F.~Becchetti, D.~Roberts,
  A.~Galonsky, R.~Ronningen, M.~Steiner, J.~Brown, J.~Kolata, A.~Nadsen,
  K.~Subotic, {Total reaction and neutron-removal cross sections of (30--60)
  $A$ MeV Be isotopes on Si and Pb}, Phys. Rev. C 64 (2001) 044611.

\bibitem{Lu13}
Z.-T. Lu, P.~Mueller, G.~F. Drake, W.~N{\"{o}}rtersh{\"{a}}user, S.~Pieper,
  Z.-C. Yan, {$Colloquium$: Laser probing of neutron-rich nuclei in light
  atoms}, Rev. Mod. Phys. 85 (2013) 1383.

\bibitem{Nortershauser09}
W.~N{\"{o}}rtersh{\"{a}}user, D.~Tiedemann, M.~{\v{Z}}{\'{a}}kov{\'{a}},
  Z.~Andjelkovic, K.~Blaum, M.~Bissell, R.~Cazan, G.~Drake, C.~Geppert,
  M.~Kowalska, J.~Kr{\"{a}}mer, A.~Krieger, R.~Neugart, R.~S{\'{a}}nchez,
  F.~Schmidt-Kaler, Z.-C. Yan, D.~Yordanov, C.~Zimmermann, {Nuclear Charge
  Radii of $^{7,9,10}$Be and the One-Neutron Halo Nucleus $^{11}$Be}, Phys.
  Rev. Lett. 102 (2009) 062503.

\bibitem{Shen96}
Y.~Shen, Z.~Ren, {Skyrme-Hartree-Fock calculation on He, Li, and Be isotopes},
  Phys. Rev. C 54 (1996) 1158.

\bibitem{Krieger12}
A.~Krieger, K.~Blaum, M.~Bissell, N.~Fr{\"{o}}mmgen, C.~Geppert, M.~Hammen,
  K.~Kreim, M.~Kowalska, J.~Kr{\"{a}}mer, T.~Neff, R.~Neugart, G.~Neyens,
  W.~N{\"{o}}rtersh{\"{a}}user, C.~Novotny, R.~S{\'{a}}nchez, D.~Yordanov,
  {Nuclear Charge Radius of $^{12}$Be}, Phys. Rev. Lett. 108 (2012) 142501.

\bibitem{Carlson15}
J.~Carlson, F.~Gandolfi, F.~Pederiva, S.~Pieper, R.~Schiavilla, K.~Schmidt,
  R.~Wiringa, {Quantum Monte Carlo methods for nuclear physics}, Rev. Mod.
  Phys. 87 (2015) 1067.

\bibitem{Adelberger11}
E.~Adelberger, A.~Garcia, R.~H. Robertson, K.~Snover, A.~Balantekin, K.~Heeger,
  M.~Ramsey-Musolf, D.~Bemmerer, A.~Junghans, C.~Bertulani, J.-W. Chen,
  H.~Costantini, P.~Prati, M.~Couder, E.~Uberseder, M.~Wiescher, R.~Cyburt,
  B.~Davids, S.~Freedman, M.~Gai, D.~Gazit, L.~Gialanella, G.~Imbriani,
  U.~Greife, M.~Hass, W.~Haxton, T.~Itahashi, K.~Kubodera, K.~Langanke,
  D.~Leitner, M.~Leitner, P.~Vetter, L.~Winslow, L.~Marcucci, T.~Motobayashi,
  A.~Mukhamedzhanov, R.~Tribble, K.~M. Nollet, F.~Nunes, T.-S. Park, P.~Parker,
  R.~Shiavilla, E.~Simpson, C.~Spitaleri, F.~Strieder, H.-P. Trautvetter,
  K.~Suemmerer, S.~Typel, {Solar fusion cross sections. II. The $pp$ chain and
  CNO cycles}, Rev. Mod. Phys. 83 (2011) 195.

\bibitem{Riisager93}
K.~Riisager, A.~Jensen, The radius of {$^8$B} and solar neutrinos, Phys. Lett.
  B 301 (1993) 6.

\bibitem{Csoto98}
A.~Cs\'{o}t\'{o}, K.~Langanke, Effects of {$^8$B} size on the low-energy
  {$^7$Be}(p,$\gamma$){$^8$B} cross section, Nucl. Phys. A 636 (1998) 240.

\bibitem{Alk78}
G.~Alkhazov, S.~Belostotsky, A.~Vorobyov, {Scattering of 1 GeV protons on
  nuclei}, Phys. Rep. 42C (1978) 89.

\bibitem{Sakaguchi17}
H.~Sakaguchi, J.~Zenihiro, {Proton elastic scattering from stable and unstable
  nuclei -- Extraction of nuclear densities}, Prog. Part. Nucl. Phys. 97 (2017)
  1.

\bibitem{Alk92}
G.~Alkhazov, A.~Lobodenko, Possibility of determining the radius of the neutron
  halo in light exotic nuclei, JETP Lett. 55 (1992) 379.

\bibitem{Alk97}
G.~Alkhazov, M.~Andronenko, A.~Dobrovolsky, P.~Egelhof, G.~Gavrilov,
  H.~Geissel, H.~Irnich, A.~Khanzadeev, G.~Korolev, A.~Lobodenko,
  G.~M{\"{u}}nzenberg, M.~Mutterer, S.~Neumaier, F.~Nickel, W.~Schwab,
  D.~Seliverstov, T.~Suzuki, J.~Theobald, N.~Timofeev, A.~Vorobyov,
  V.~Yatsoura, {Nuclear Matter Distributions in $^6$He and $^8$He from Small
  Angle p-He Scattering in Inverse Kinematics at Intermediate Energy}, Phys.
  Rev. Lett. 78 (1997) 2313.

\bibitem{Alk02}
G.~Alkhazov, A.~Dobrovolsky, P.~Egelhof, H.~Geissel, H.~Irnich, A.~Khanzadeev,
  G.~Korolev, A.~Lobodenko, G.~M{\"{u}}nzenberg, M.~Mutterer, S.~Neumaier,
  W.~Schwab, D.~Seliverstov, T.~Suzuki, A.~Vorobyov, {Nuclear matter
  distributions in the $^6$He and $^8$He nuclei from differential cross
  sections for small-angle proton elastic scattering at intermediate energy},
  Nucl. Phys. A 712 (2002) 269.

\bibitem{Neumaier02}
S.~Neumaier, G.~Alkhazov, M.~Andronenko, A.~Dobrovolsky, P.~Egelhof,
  G.~Gavrilov, H.~Geissel, H.~Irnich, A.~Khanzadeev, G.~Korolev, A.~Lobodenko,
  G.~M{\"{u}}nzenberg, M.~Mutterer, W.~Schwab, D.~Seliverstov, T.~Suzuki,
  N.~Timofeev, A.~Vorobyov, V.~Yatsoura, {Small-angle proton elastic scattering
  from the neutron-rich isotopes $^6$He and $^8$He, and from $^4$He, at 0.7 GeV
  in inverse kinematics}, Nucl. Phys. A 712 (2002) 247.

\bibitem{Egelhof02}
P.~Egelhof, G.~Alkhazov, M.~Andronenko, A.~Bauchet, A.~Dobrovolsky, S.~Fritz,
  G.~Gavrilov, H.~Geissel, C.~Gross, A.~Khanzadeev, G.~Korolev, G.~Kraus,
  A.~Lobodenko, G.~M{\"{u}}nzenberg, M.~Mutterer, S.~Neumaier,
  T.~Sch{\"{a}}fer, C.~Scheidenberger, D.~Seliverstov, N.~Timofeev,
  A.~Vorobyov, V.~Yatsoura, Nuclear-matter distributions of halo nuclei from
  elastic proton scattering in inverse kinematics, Eur. Phys. J. A 15 (2002)
  27.

\bibitem{Dobrov06}
A.~Dobrovolsky, G.~Alkhazov, M.~Andronenko, A.~Bauchet, P.~Egelhof, S.~Fritz,
  H.~Geissel, C.~Gross, A.~Khanzadeev, G.~Korolev, G.~Kraus, A.~Lobodenko,
  G.~M{\"{u}}nzenberg, M.~Mutterer, S.~Neumaier, T.~Sch{\"{a}}fer,
  C.~Scheidenberger, D.~Seliverstov, N.~Timofeev, A.~Vorobyov, V.~Yatsoura,
  {Study of the nuclear matter distribution in neutron-rich Li isotopes}, Nucl.
  Phys. A 766 (2006) 1.

\bibitem{Ilieva12}
S.~Ilieva, F.~Aksouh, G.~Alkhazov, L.~Chulkov, A.~Dobrovolsky, P.~Egelhof,
  H.~Geissel, M.~Gorska, A.~Inglessi, R.~Kanungo, A.~Khanzadeev, O.~Kiselev,
  G.~Korolev, X.~Le, Y.~Litvinov, C.~Nociforo, D.~Seliverstov, L.~Sergeev,
  H.~Simon, V.~Volkov, A.~Vorobyov, H.~Weick, V.~Yatsoura, A.~Zhdanov,
  {Nuclear-matter density distribution in the neutron-rich nuclei $^{12,14}$Be
  from proton elastic scattering in inverse kinematics}, Nucl. Phys. A 875
  (2012) 8.

\bibitem{Korolev18}
G.~Korolev, A.~Dobrovolsky, A.~Inglessi, G.~Alkhazov, P.~Egelhof,
  A.~Estrad\'{e}, I.~Dillmann, F.~Farinon, H.~Geissel, S.~Ilieva, Y.~Ke,
  A.~Khanzadeev, O.~Kiselev, J.~Kurcewicz, X.~Le, Y.~Litvinov, G.~Petrov,
  A.~Prochazka, C.~Scheidenberger, L.~Sergeev, H.~Simon, M.~Takechi, S.~Tang,
  V.~Volkov, A.~Vorobyov, H.~Weick, V.~Yatsoura, Halo structure of {$^8$B}
  determined from intermediate energy proton elastic scattering in inverse
  kinematics, Phys. Letters B 780 (2018) 200.

\bibitem{Vor74}
A.~Vorobyov, G.~Korolev, V.~Schegelsky, G.~Solyakin, G.~Sokolov, Y.~Zalite, A
  method for studies of small-angle hadron-proton elastic scattering in the
  coulomb interference region, Nucl. Instr. Meth. 119 (1974) 509.

\bibitem{Vor82}
A.~Vorobyov, Y.~Grigorev, Y.~Zalite, G.~Korolev, E.~Maev, G.~Sokolov,
  A.~Khanzadeev, An ionization spectrometer for recoil nuclei in research on
  elastic small-angle scattering of hadrons, Instrum. Exp. Tech. 24 (1982)
  1127.

\bibitem{Burq78}
J.~Burq, M.~Chemarin, M.~Chevallier, A.~Denisov, C.~Dor{\'{e}},
  T.~Ekel{\"{o}}f, P.~Grafstr{\"{o}}m, E.~Hagberg, B.~Ille, A.~Kashchuk,
  G.~Korolev, S.~Kullander, M.~Lambert, J.~Martin, S.~Maury, J.~Paumier,
  M.~Querrou, V.~Schegelsky, E.~Spiridenkov, I.~Tkach, A.~Vorobyov,
  {Measurements of $\pi^-$p elastic scattering in the coulomb interference
  region at high energies}, Phys. Lett. B 77 (1978) 438.

\bibitem{Burq83}
J.~Burq, M.~Chemarin, M.~Chevallier, A.~Denisov, C.~Dor{\'{e}},
  T.~Ekel{\"{o}}f, J.~Fay, P.~Grafstr{\"{o}}m, L.~Gustafsson, E.~Hagberg,
  B.~Ille, A.~Kashchuk, G.~Korolev, A.~Kulikov, S.~Kullander, M.~Lambert,
  J.~Martin, S.~Maury, M.~Querrou, V.~Schegelsky, E.~Spiridenkov, I.~Tkach,
  M.~Verbeken, A.~Vorobyov, {Soft $\pi^-$p and pp elastic scattering in the
  energy range 30 to 345 GeV}, Nucl. Phys. B 217 (1983) 285.

\bibitem{Tassie56}
L.~Tassie, {A Model of Nuclear Shape Oscillations for $\gamma$-Transitions and
  Electron Excitation}, Austr. J. Phys. 9 (1956) 407.

\bibitem{Colo18}
G.~Col\`{o}, {\it {to be published}}.

\bibitem{Carstoiu01}
F.~Carstoiu, L.~Trache, C.~Gagliardi, R.~Tribble, A.~Mukhamedzhanov, {Radius of
  $^8$B halo from the asymptotic normalization coefficient}, Phys. Rev. C 63
  (2001) 054310.

\bibitem{Liu04}
Z.-H. Liu, J.-D. Bao, {P-Wave Nuclear Halos in $^8$B and $^{11}$Be}, Chin.
  Phys. Lett. 21 (2004) 457.

\bibitem{Carstoiu07}
F.~Carstoiu, L.~Trache, C.~Gagliardi, A.~Mukhamedzhanov, R.~Tribble, {Radii of
  halo states in light nuclei deduced from ANC}, Roman. Rep. Phys. 59 (2007)
  357.

\bibitem{Angeli13}
I.~Angeli, K.~Marinova, {Table of experimental nuclear ground state charge
  radii: An update}, At. Data Nucl. Data Tables 99 (2013) 69.

\bibitem{Estrade14}
A.~Estrad{\'{e}}, R.~Kanungo, W.~Horiuchi, F.~Ameil, J.~Atkinson, Y.~Ayyad,
  D.~Cortina-Gil, I.~Dillmann, A.~Evdokimov, F.~Farinon, H.~Geissel,
  G.~Guastalla, R.~Janik, M.~Kimura, R.~Kn{\"{o}}bel, J.~Kurcewicz,
  Y.~Litvinov, M.~Marta, M.~Mostazo, I.~Mukha, C.~Nociforo, H.~Ong, S.~Pietri,
  A.~Prochazka, C.~Scheidenberger, B.~Sitar, P.~Strmen, Y.~Suzuki, M.~Takechi,
  J.~Tanaka, I.~Tanihata, S.~Terashima, J.~Vargas, H.~Weick, J.~Winfield,
  {Proton Radii of $^{12-17}$B Define a Thick Neutron Surface in $^{17}$B},
  Phys. Rev. Lett. 113 (2014) 132501.

\end{thebibliography}

\clearpage
\section*{Appendix}
This Appendix contains in tabular form the cross sections d$\sigma$/d$t$ as a function of the four-momentum transfer squared for $p^8$B and  $p^7$Be elastic scattering measured in the present experiment. Only statistical errors are indicated.

\begin{tabular}{ccccc}
	\multicolumn{4}{c}{}                            &  \\ \hline
	\multicolumn{2}{c}{$p^{8}$B, $E_{\rm p}$=701.8 MeV}  &  &  \multicolumn{2}{c}{$p^{8}$B, $E_{\rm p}$=701.8 MeV}  \\ \hline
	$-t$, (GeV/$c)^2$ & d$\sigma$/d$t$,  mb/(GeV/$c)^2$ &  & $-t$, (GeV/$c)^2$ & d$\sigma$/d$t$,  mb/(GeV/$c)^2$ \\ \hline
	0.00117      &         9060 $\pm$ 207          &  &      0.01548      &          1294 $\pm$ 47          \\
	0.00164      &         5876 $\pm$ 125          &  &      0.01684      &          1152 $\pm$ 43          \\
	0.00211      &         4776 $\pm$ 133          &  &      0.01826      &          1002 $\pm$ 40          \\
	0.00258      &         4043 $\pm$ 123          &  &      0.01973      &        946.9 $\pm$ 38.2         \\
	0.00305      &         3555 $\pm$ 124          &  &      0.02127      &        865.5 $\pm$ 36.0         \\
	0.00352      &         3343 $\pm$ 119          &  &      0.02285      &        791.9 $\pm$ 34.0         \\
	0.00399      &         3077 $\pm$ 123          &  &      0.02450      &        654.2 $\pm$ 30.5         \\
	0.00446      &         3031 $\pm$ 113          &  &      0.02620      &        620.9 $\pm$ 29.3         \\
	0.00493      &         2765 $\pm$ 113          &  &      0.02796      &        551.2 $\pm$ 27.3         \\
	0.00540      &         2586 $\pm$ 114          &  &      0.02978      &        492.6 $\pm$ 25.5         \\
	0.00586      &         2536 $\pm$ 104          &  &      0.03165      &        438.9 $\pm$ 23.8         \\
	0.00633      &         2348 $\pm$  99          &  &      0.03359      &        403.7 $\pm$ 22.6         \\
	0.00680      &         2563 $\pm$ 104          &  &      0.03558      &        321.7 $\pm$ 20.0         \\
	0.00727      &         2247 $\pm$  99          &  &      0.03762      &        314.1 $\pm$ 19.6         \\
	0.00774      &         2138 $\pm$  97          &  &      0.03973      &        273.4 $\pm$ 18.1         \\
	0.00852      &         2149 $\pm$  68          &  &      0.04189      &        211.3 $\pm$ 15.7         \\
	0.00954      &         1891 $\pm$  62          &  &      0.04411      &        207.3 $\pm$ 15.5         \\
	0.01061      &         1853 $\pm$  60          &  &      0.04755      &         160.0 $\pm$ 9.5         \\
	0.01174      &         1622 $\pm$  55          &  &      0.05233      &         111.8 $\pm$ 7.8         \\
	0.01293      &         1572 $\pm$  53          &  &      0.05735      &         89.7 $\pm$ 6.9          \\
	0.01418      &         1296 $\pm$  48          &  &                   &  \\ \hline
\end{tabular}
\newpage
\begin{tabular}{ccccc}
	\hline
	\multicolumn{2}{c}{$p^7$Be, $E_{\rm p}$ = 701.1 MeV}  &  & \multicolumn{2}{c}{$p^7$Be, $E_{\rm p}$ = 701.1 MeV}  \\ \hline
	$-t$, (GeV/$c)^2$ & d$\sigma$/d$t$,  mb/(GeV/$c)^2$ &  & $-t$, (GeV/$c)^2$ & d$\sigma$/d$t$,  mb/(GeV/$c)^2$ \\ \hline
	0.00117      &         6401 $\pm $ 171         &  &      0.01739      &        921.6 $\pm $ 41.5        \\
	0.00164      &         4317 $\pm $ 138         &  &      0.01864      &        931.5 $\pm $ 43.0        \\
	0.00211      &         3749 $\pm $ 127         &  &      0.01993      &        773.4 $\pm $ 39.6        \\
	0.00258      &         3093 $\pm $ 115         &  &      0.02127      &        775.1 $\pm $ 39.2        \\
	0.00305      &         2795 $\pm $ 109         &  &      0.02265      &        664.4 $\pm $ 36.0        \\
	0.00352      &         2519 $\pm $ 103         &  &      0.02407      &        666.2 $\pm $ 35.0        \\
	0.00399      &         2519 $\pm $ 102         &  &      0.02553      &        592.5 $\pm $ 33.4        \\
	0.00446      &         2338 $\pm $ 98          &  &      0.02704      &        567.2 $\pm $ 32.4        \\
	0.00493      &         2191 $\pm $ 95          &  &      0.02860      &        522.8 $\pm $ 31.4        \\
	0.00540      &         1962 $\pm $ 90          &  &      0.03019      &        451.8 $\pm $ 28.6        \\
	0.00586      &         2011 $\pm $ 92          &  &      0.03183      &        416.0 $\pm $ 27.3        \\
	0.00633      &         2061 $\pm $ 92          &  &      0.03351      &        372.5 $\pm $ 23.6        \\
	0.00680      &         1889 $\pm $ 88          &  &      0.03612      &        298.3 $\pm $ 14.7        \\
	0.00728      &         1873 $\pm $ 76          &  &      0.03975      &        248.8 $\pm $ 14.0        \\
	0.00810      &         1692 $\pm $ 69          &  &      0.04355      &        180.5 $\pm $ 11.1        \\
	0.00896      &         1605 $\pm $ 68          &  &      0.04752      &        161.8 $\pm $ 11.6        \\
	0.00986      &         1546 $\pm $ 64          &  &      0.05168      &        120.0 $\pm $ 9.5         \\
	0.01081      &         1510 $\pm $ 62          &  &      0.05600      &         96.7 $\pm $ 8.8         \\
	0.01180      &         1335 $\pm $ 59          &  &      0.06051      &         78.3 $\pm $ 7.7         \\
	0.01283      &         1350 $\pm $ 58          &  &      0.06519      &         51.1 $\pm $ 6.3         \\
	0.01391      &         1092 $\pm $ 47          &  &      0.07005      &         32.3 $\pm $ 5.1         \\
	0.01503      &         1127 $\pm $ 52          &  &      0.07508      &         31.7 $\pm $ 4.4         \\
	0.01619      &         1020 $\pm $ 47          &  &      0.08029      &         21.8 $\pm $ 3.6         \\
	0.01919      &         1025 $\pm $ 42          &  &                   &                                 \\ \hline
\end{tabular}

\end{document}